\documentclass{article}
\usepackage{spconf,amsmath,graphicx,hyperref}
\usepackage{algorithm}
\usepackage{algpseudocode}
\usepackage{tabularx}
\usepackage{booktabs}
\usepackage{xcolor}
\usepackage{multirow, caption}
\usepackage[utf8]{inputenc}

\title{LongSpeech: A Scalable Benchmark for Transcription, Translation and Understanding in Long Speech}

\name{
\begin{tabular}[t]{c}
     Fei Yang\textsuperscript{1,2},
  Xuanfan Ni\textsuperscript{1},
  Renyi Yang\textsuperscript{1,3}\thanks{This work was done during Fei Yang and Renyi Yang's internships at Alibaba International Digital Commerce.}, Jiahui Geng\textsuperscript{4}, Qing Li\textsuperscript{5}, 
  Chenyang Lyu\textsuperscript{1,*}\thanks{*Corresponding author.}, \\ Yichao Du\textsuperscript{1},
  Longyue Wang\textsuperscript{1}, 
  \textit{Weihua Luo}\textsuperscript{1},
  \textit{Kaifu Zhang}\textsuperscript{1}
\end{tabular}
}

\address{$^{1}$Alibaba International Digital Commerce  \hspace{1em} $^{2}$Shanghai Jiao Tong University \\ 
         $^{3}$Delft University of Technology       \hspace{1em}\textsuperscript{4}Linköping University\hspace{1em}\textsuperscript{5}University of Groningen}

\begin{document}
%
\maketitle
\begin{abstract}

Recent advances in audio-language models have demonstrated remarkable success on short, segment-level speech tasks. However, real-world applications such as meeting transcription, spoken document understanding, and conversational analysis require robust models capable of processing and reasoning over long-form audio. In this work, we present \textbf{LongSpeech}, a large-scale and scalable benchmark specifically designed to evaluate and advance the capabilities of speech models on long-duration audio. LongSpeech comprises over 100,000 speech segments, each approximately 10 minutes long, with rich annotations for ASR, speech translation, summarization, language detection, speaker counting, content separation, and question answering. We introduce a reproducible pipeline for constructing long-form speech benchmarks from diverse sources, enabling future extensions. Our initial experiments with state-of-the-art models reveal significant performance gaps, with models often specializing in one task at the expense of others and struggling with higher-level reasoning. These findings underscore the challenging nature of our benchmark. Our benchmark will be made publicly available to the research community.

\end{abstract}
\section{Introduction}
\label{sec:intro}

The fields of automatic speech recognition (ASR) and audio-language modeling have achieved significant milestones, particularly on tasks involving short or moderately long audio segments \cite{radford2022robust_whisper}. State-of-the-art models now routinely attain near-human performance on standard benchmarks like LibriSpeech \cite{panayotov2015librispeech} and TED-LIUM \cite{Hernandez_2018_TED_LIUM}, which predominantly focus on isolated utterances or brief exchanges. These advances have been accelerated by the emergence of audio language models \cite{huang2025step,chu2024qwen2audiotechnicalreport,kimiteam2025kimiaudiotechnicalreport,li2025baichuan}, which enable joint speech understanding and language generation. Despite these breakthroughs, a persistent gap remains between these controlled benchmarks and the demands of real-world applications, where audio often extends over many minutes or hours and contains complex phenomena such as speaker changes, topic shifts, and multilingual content.

\textbf{LongSpeech} is designed to bridge this gap by providing a comprehensive, large-scale benchmark tailored for long-form speech. Processing long-duration audio introduces unique challenges, including maintaining semantic coherence, handling speaker and language variability, and supporting complex downstream tasks. Existing benchmarks do not sufficiently capture these complexities. Although recent work has introduced long-form evaluations such as BLAB \cite{ahia2025blab}, which focus on temporal localization and counting, they still lack support for higher-level semantic and discourse-level understanding. This gap in comprehensive evaluation capabilities limits our ability to develop and objectively assess models for real-world deployment. Our work focuses on two core contributions: a scalable framework for constructing diverse, long-form speech datasets and a multi-task benchmark designed to rigorously test advanced audio-language models.

To this end, LongSpeech aggregates long-duration audio from diverse sources and provides annotations for a variety of tasks that reflect practical needs. These tasks are intentionally challenging and range from fundamental speech perception to higher-level reasoning:

\begin{itemize}\itemsep0pt
\item \textbf{ASR \& S2T Translation:} Core transcription and translation of full-length audio.
\item \textbf{Summarization:} Generating concise summaries from lengthy recordings.
\item \textbf{Speaker Count \& Language Detection:} Identifying speaker and language attributes.
\item \textbf{Content Separation:} Detecting unrelated concatenated content to test coherence.
\item \textbf{QA \& Temporal Localization:} Evaluating comprehension, reasoning, and temporal tracking.
\item \textbf{Emotion Analysis:} Determining the overall emotional tone of the speech.
\end{itemize}

By encompassing this diverse set of tasks, LongSpeech provides a unified platform to benchmark not only the transcription and translation accuracy of models but also their ability to understand and reason about long-form audio. As our initial experiments demonstrate, even leading models exhibit significant weaknesses on these tasks, underscoring the difficulty of the benchmark and the urgent need for progress. We believe that LongSpeech will catalyze research towards more robust and versatile audio-language models, ultimately closing the gap between academic benchmarks and real-world utility.

\section{Dataset Construction and Curation}
\label{sec:dataset}

The LongSpeech benchmark is meticulously constructed to provide a diverse, large-scale resource for evaluating long-form speech processing. The dataset consists of over 100,000 long speech segments, each with a typical duration of approximately 10 minutes. The construction and curation process integrates speech data from a wide array of publicly available corpora, as well as carefully synthesized content, ensuring broad representation in terms of domains, speakers, and languages.

\subsection{Data Sources}

LongSpeech draws from multiple high-quality speech datasets, including LibriSpeech, TED-LIUM v3, SPGISpeech, VoxPopuli, CommonVoice, AISHELL-2, IWSLT, and a custom movie dialogue corpus. These sources cover various genres and languages, with licenses that permit research use. The composition of LongSpeech is as follows:

\begin{table}[htbp]
\centering
\caption{Datasets used in benchmark construction.}
\label{tab:datasets_resized}
\resizebox{\linewidth}{!}{%
\begin{tabular}{lll}
\toprule
\textbf{Dataset} & \textbf{Description} & \textbf{Proportion / Size} \\
\midrule
\textbf{LibriSpeech}   & English audiobook readings                  & 6\% \\
\textbf{TED-LIUM v3}   & English TED talks                           & 4\% \\
\textbf{SPGISpeech}    & Financial phone conversations               & 30\% \\
\textbf{VoxPopuli}     & Multilingual parliamentary meetings         & 20\% \\
\textbf{CommonVoice}   & Crowdsourced multilingual short utterances & 34\% \\
\textbf{AISHELL-2}     & Mandarin Chinese speech                     & 6\% \\
\textbf{IWSLT}         & TED talks                                   & 50 records \\
\textbf{Movie-corpus}  & Synthesized multi-speaker conversations    & 350 records \\
\bottomrule
\end{tabular}%
}
\end{table}

\begin{table*}[ht!]
    \centering
    \caption{LongSpeech Task Partitioning: total, train, dev, and test set sizes per task.}
    \resizebox{\linewidth}{!}{
    \begin{tabular}{|l|p{6cm}|p{4cm}|r|r|r|r|}
        \hline
        \textbf{Task} & \textbf{Sample Question (Q)} & \textbf{Sample Answer (A)} & \textbf{Total} & \textbf{Train} & \textbf{Dev} & \textbf{Test} \\
        \hline
        S2TT (Speech-to-Text Translation) & Detect the language and translate the speech into French & Venant comme cela à une période.. & 42k & 29k & 6.3k & 6.3k \\
        \hline
        ASR (Automatic Speech Recognition) & Detect the language and recognize the speech: \texttt{<|en|>} & A quick brown fox jumps over a lazy dog & 102k & 71k & 15k & 15k \\
        \hline
        Summarization & Provide a summary of the main topic discussed & The team discussed the Q3 budget shortfall... & 6.2k & 4.3k & 0.9k & 0.9k \\
        \hline
        Speaker Count & How many unique speakers are present & 4 & 8.4k & 5.8k & 1.2k & 1.2k \\
        \hline
        Language Detection & What is the language used during the speech & English & 21.1k & 14.7k & 3.1k & 3.1k \\
        \hline
        Content Separation & How many unrelated segments are there in the audio? & 2 & 8.4k & 5.8k & 1.2k & 1.2k \\
        \hline
        Emotion Analysis & What's the emotional vibe of this audio? & Suspenseful & 8k & 5.8k & 1.2k & 1.2k \\
        \hline
        Temporal Issue Localization & Regarding the speech's progression, what 10th topic was presented? & the interpretation of the dream & 8k & 5.8k & 1.2k & 1.2k \\
        \hline
    \end{tabular}
    }
    \label{tab:tasks}
\end{table*}

\subsection{Curation Methodology}

To assemble long-duration segments, we employ data-specific strategies:

\begin{itemize}\itemsep0pt
    \item \textbf{Speaker and Topic Coherence}: For LibriSpeech and SPGISpeech, utterances are grouped by speaker and chapter, concatenating sequentially to reach $\sim 600$\,s duration.
    \item \textbf{Embedding-Based Selection}: For CommonVoice, sentence embeddings and FAISS-based clustering group semantically similar segments. Speaker embeddings ensure diversity across concatenated content.
    \item \textbf{Multilingual Processing}: VoxPopuli and AISHELL-2 prioritize supervised, multi-speaker segments while excluding very short utterances.
    \item \textbf{Synthetic Content}: Movie-corpus uses text-to-speech with diverse speaker and gender distributions.
    \item \textbf{Task-Specific Curation}: LibriSpeech and TED-LIUM segments are selected for summarization and QA tasks based on content quality and complexity.
\end{itemize}

\subsection{Annotation and Task Preparation}

Each long-form audio segment is further annotated for multiple downstream tasks, including ASR, translation, summarization, language detection, speaker counting, content separation, and spoken question answering (QA). Ground-truth transcriptions are sourced from original datasets or generated with high-quality models where human annotation is absent. For tasks such as speaker count, language detection, and content separation, metadata is extracted or inferred from corpus-level annotations.

LongSpeech encompasses a multi-task benchmark, annotated for eight distinct tasks designed to evaluate both low-level and high-level speech understanding. Each task is partitioned into train,  dev, and test splits, facilitating robust training and evaluation protocols. All tasks share a unified directory structure, with train, validation, and test splits distributed in a 7:1.5:1.5 ratio. All splits within each task are combined respectively to form the final unified train, dev, and test sets for the LongSpeech benchmark. In total, the LongSpeech benchmark comprises 142,200 examples in the training set, 30,100 examples in the dev set, and 30,100 examples in the test set. These totals are obtained by aggregating the respective splits from all eight tasks, ensuring that each split is both comprehensive and representative of the dataset’s multi-task nature.

\section{Experiments on LongSpeech}

\subsection{Baseline Systems}

To facilitate comparative evaluation, several strong foundation audio-language models are used in experiments, including Qwen2Audio \cite{chu2024qwen2audiotechnicalreport}, KimiAudio \cite{kimiteam2025kimiaudiotechnicalreport}, AudioFlamingo3 \cite{goel2025audioflamingo3advancing}, Voxtral \cite{liu2025voxtral} and DashengLM \cite{dinkel2025midashenglm}.

\subsection{Evaluation Metrics}
\label{sec:metrics}

We describe the evaluation metrics used for each task, organized by functionality and output type. \textbf{Automatic Speech Recognition}
We evaluate speech recognition performance using: {Word Error Rate (WER)}: the ratio of word-level edit errors (insertions, deletions, substitutions) to total words in the reference transcription. Lower values indicate better performance. \textbf{Speech-to-Text Translation}
For end-to-end translation from speech to text, we use {BLEU}~\cite{papineni-etal-2002-bleu}: n-gram precision with brevity penalty, computed case-insensitively as BLEU-4. Higher scores indicate better translation fluency and adequacy. \textbf{Summarization} We assess summary quality using recall-oriented n-gram overlap metrics {ROUGE}~\cite{lin-2004-rouge}: we report ROUGE-1 (unigram), ROUGE-2 (bigram), and ROUGE-L (longest common subsequence) F1 scores. Higher values indicate better content coverage and coherence.

\subsubsection{Fixed-Answer Tasks}
This category includes \textit{Content Separation} and \textit{Speaker Count}, where models must extract or generate structured, discrete answers (e.g., numbers, counts). Both tasks share the same evaluation protocol: 1) \textbf{Numeric Accuracy}: proportion of responses that exactly match the ground truth numeric value. 2) \textbf{Parsability Rate}: fraction of inputs where the model successfully parses the query structure (e.g., identifies target entity and operation). 3) \textbf{Post-Parsing Precision}: accuracy of answers among successfully parsed queries. 4) \textbf{Misunderstanding Rate}: proportion of cases where the model misinterprets the question intent (e.g., confuses entities or operations).
These metrics jointly reflect both parsing capability and answer correctness.

\begin{table}[t]
\centering
\caption{Speech Recognition and Translation Performance}
\label{tab:asr_s2tt_simple}
\small
\setlength{\tabcolsep}{5pt}
\renewcommand{\arraystretch}{1.15}
\resizebox{\linewidth}{!}{
\begin{tabular}{l c c c c}
\toprule
\textbf{Model} 
& \textbf{Non-CJK WER $\downarrow$} 
& \textbf{CJK CER $\downarrow$} 
& \textbf{Overall CER $\downarrow$}
& \textbf{S2TT BLEU $\uparrow$} \\
\midrule
Whisper\cite{radford2022robust_whisper}         & 0.186 & 0.385 & 0.110 & —— \\
Kimi-audio\cite{kimiteam2025kimiaudiotechnicalreport}      & 0.542 & 0.905 & 0.501 & 15.81 \\
AudioFlamingo3\cite{goel2025audioflamingo3advancing}  & 1.378 & 1.501 & 1.595 & 0.03 \\
Voxtral\cite{liu2025voxtral}         & 0.228 & 0.849 & 0.188 & 30.20 \\
DashengLM\cite{dinkel2025midashenglm}       & 0.389 & 0.759 & 0.311 & 5.48 \\
Qwen2-Audio\cite{chu2024qwen2audiotechnicalreport}     & 0.298 & 0.709 & 0.253 & 11.39 \\
\bottomrule
\end{tabular}
}
\footnotesize
\textit{Note:} 
Non-CJK includes English, French, German, Spanish, etc.; 
CJK includes Japanese, Korean, and Chinese.
\end{table}

\begin{table*}[t]
\centering
\caption{Performance Comparison on Comprehensive Audio Understanding Tasks including Content Separation, Emo. (Emotion Analysis), Speaker Count, Summary, Lang. Det. (Language Detection), and Temp. Loc. (Temporal Issue Localization). Metric abbreviations are as follows: N.A (Numeric Accuracy), P.R (Parsability Rate), Pr (Post-Parsing Precision), M.R (Misunderstanding Rate), St.A (Strict Accuracy), R.A (Relaxed Accuracy), R1 (ROUGE-1 F1), R2 (ROUGE-2 F1), RL (ROUGE-L F1), D.A (Detection Accuracy), and D.E (Detection Errors).}

\label{tab:understanding_full_names}
\small
\resizebox{\textwidth}{!}{%
\setlength{\tabcolsep}{6pt}
\renewcommand{\arraystretch}{1.25}
\begin{tabular}{l c c c c c c c c c c c c c c c c c c}
\toprule
\textbf{Model} & 
\multicolumn{4}{c}{\textbf{Content Separation}} & 
\multicolumn{2}{c}{\textbf{Emo.}} & 
\multicolumn{4}{c}{\textbf{Speaker Count}} & 
\multicolumn{3}{c}{\textbf{Summary}} & 
\multicolumn{3}{c}{\textbf{Lang. Det.}} & 
\multicolumn{2}{c}{\textbf{Temp. Loc.}} \\
\cmidrule(lr){2-5} \cmidrule(lr){6-7} \cmidrule(lr){8-11} \cmidrule(lr){12-14} \cmidrule(lr){15-17} \cmidrule(lr){18-19}
& {N.A} & {P.R} & {Pr} & {M.R} & {St.A} & {R.A} & {N.A} & {P.R} & {Pr} & {M.R} & {R1} & {R2} & {RL} & {D.A} & {M.R} & {D.E} & {St.A} & {R.A} \\
\midrule
AudioFlamingo3\cite{goel2025audioflamingo3advancing} & 3.33 & 66.27 & 5.03 & 33.73 & 18.53 & 34.13 & 21.62 & 70.63 & 30.61 & 29.37 & 20.25 & 4.92 & 12.97 & 88.29 & 9.85 & 1.86 & 6.10 & 12.52 \\
Voxtral\cite{liu2025voxtral} & 25.73 & 76.01 & 33.85 & 23.99 & 38.80 & 51.62 & 28.50 & 99.76 & 28.57 & 0.24 & 41.81 & 14.61 & 25.10 & 79.24 & 20.38 & 0.38 & 23.69 & 48.81 \\
DashengLM\cite{dinkel2025midashenglm} & 23.75 & 99.05 & 23.98 & 0.95 & 11.08 & 29.53 & 35.31 & 100 & 35.31 & 0 & 15.22 & 1.24 & 10.38 & 9.37 & 0 & 90.63 & 0.48 & 6.10 \\
\bottomrule
\end{tabular}%
}
\end{table*}

\subsubsection{Emotion Analysis}
Predictions are mapped from fine-grained labels into seven coarse emotion categories:
We group fine-grained emotion predictions into seven coarse categories: 1)  \textit{Positive-HighArousal}: Excited, Amused. 2) \textit{Positive-Uplifting}: Hopeful, Determined, Inspirational, Awe, Vivid. 3) \textit{Negative-Sadness/Reflection}: Melancholy, Somber, Pensive, Reflective. 4) \textit{Negative-Tension/Seriousness}: Suspenseful, Serious, Dramatic, Urgent, Intense. 5) \textit{Cognitive-Curiosity}: Intrigued, Questioning. 6) \textit{Cognitive-Uncertainty}: Confused 7) \textit{Neutral-Informative}: Neutral, Calm, Objective, Informative, Assertive.

We report \textbf{Strict Accuracy}: percentage of predictions matching the coarse label exactly and \textbf{Relaxed Accuracy}: rate at which the predicted emotion shares the same broad sentiment polarity (positive, negative, cognitive, neutral), even if the subtype differs.

\subsubsection{Temporal Issue Localization}
Responses are evaluated by GPT-4-Turbo~\cite{openai2023gpt4} using three judgment types: \textit{YES} (fully correct), \textit{NO} (incorrect), or \textit{PARTIALLY} (partially correct). Manual inspection confirms high alignment with human judgments. We use \textbf{Strict Accuracy}: ratio of \textit{YES} judgments and \textbf{Relaxed Accuracy}: ratio of \textit{YES} or \textit{PARTIALLY} judgments.

\subsection{Experimental Results and Analysis}
\label{sec:results}
Our experimental evaluation on the LongSpeech benchmark reveals that even state-of-the-art audio-language models exhibit significant performance limitations, underscoring the benchmark's challenging nature and highlighting key gaps in current long-form speech processing capabilities. 
Among the evaluated audio-language models, \textbf{AudioFlamingo3}, \textbf{Voxtral}, and \textbf{DashengLM} natively support long-audio processing. However, DashengLM cannot perform ASR or S2TT on complete long-form speech and would only output strings such as "[music]"; therefore, a segmentation-based approach was applied for both its ASR and S2TT evaluations.

\subsubsection{Specialization and Trade-Offs in Core Speech Tasks}

The results for ASR and S2TT, presented in Table \ref{tab:asr_s2tt_simple}, demonstrate a clear specialization among models rather than well-rounded performance. For reference, we also include results from \textbf{Whisper}, a specialized ASR model that does not belong to the audio-language model category and does not support direct S2TT. \textbf{Voxtral} delivers exceptional performance in speech-to-text translation, achieving the highest BLEU score of 30.20, which underscores its strong cross-lingual transfer capabilities. \textbf{Qwen2-Audio} achieves a moderate overall CER of 0.253 and a BLEU score of 11.39, reflecting relatively balanced but not top-tier performance in both tasks. Other models like \textbf{Kimi-audio} and \textbf{DashengLM} exhibit weaker results, with Kimi-audio achieving a BLEU score of 15.81 but a high overall CER of 0.501, while DashengLM obtains modest results on both tasks (CER of 0.311 and BLEU of 5.48). \textbf{AudioFlamingo3} performs poorly across both ASR and S2TT, with an overall CER of 1.595 and a near-zero BLEU score of 0.03, illustrating that large-scale generic models may not be directly suitable for long-form speech-to-text tasks without specific adaptations.

\subsubsection{Deficiencies in Higher-Level Understanding Tasks}

The limitations of current models are more apparent in the higher-level understanding tasks, as shown in Table \ref{tab:understanding_full_names}. Across all evaluated tasks, model performance is far from sufficient, indicating that robust reasoning over long audio contexts remains an open problem. A main observation is the discrepancy between understanding a query and providing an accurate answer. For instance, in \textbf{Speaker Count}, Voxtral correctly parses the user's intent with a 99.76\% Parsability Rate, yet its \textbf{Numeric Accuracy} is only 28.50\%. This issue is even more pronounced for \textbf{DashengLM}, which, in Speaker Count, achieves a perfect 100\% Parsability Rate and a 0\% Misunderstanding Rate, while its Numeric Accuracy is only 35.31\%. A similar pattern is observed in \textbf{Content Separation}. This suggests that while models can identify what is being asked, they lack the ability to analyze the audio content with the precision needed to extract the correct information.

Performance on tasks requiring deep semantic reasoning is particularly low. In \textbf{Summarization}, Voxtral's ROUGE-1 score of 41.81 indicates it can identify relevant keywords, but its low ROUGE-2 (14.61) and ROUGE-L (25.10) scores suggest difficulty in generating coherent, structurally sound summaries. The performance of \textbf{DashengLM} is significantly weaker, achieving a ROUGE-1 score of just 15.22. The results are even more notable for \textbf{Temporal Issue Localization}, the most complex reasoning task. Voxtral's strict accuracy of just 23.69\% demonstrates that tracking the progression of topics within a 10-minute recording is a difficult challenge that current models are not yet able to solve effectively. This difficulty is further underscored by \textbf{DashengLM}, which achieves a strict accuracy of only 0.48\% on the same task. Finally, tasks requiring fine-grained analysis, such as \textbf{Emotion Analysis}, also prove difficult, with the best strict accuracy falling below 40\%. This highlights the challenge of capturing subtle prosodic and contextual cues over extended and potentially multi-speaker audio segments. In summary, our findings validate the difficulty and necessity of the LongSpeech benchmark.

\section{Conclusion}

In this work, we introduced LongSpeech, a large-scale, multi-task benchmark designed to evaluate and advance long-form speech understanding. Our evaluation of state-of-the-art audio-language models reveals significant limitations in processing long-duration audio. We observed a clear trade-off between core tasks like ASR and speech translation, and a more obvious deficiency in higher-level reasoning tasks such as summarization and temporal localization. These results validate the challenging nature of LongSpeech and highlight critical gaps in current models' ability to maintain context, perform structured analysis, and reason over extended audio streams. By providing a comprehensive and scalable evaluation platform, we aim for LongSpeech to serve as a suitable testbed for long-form speech processing for advanced audio-language models.

\bibliographystyle{IEEEbib}
\bibliography{strings,refs}

\end{document}